# Recent Advances in Wearable Sensors with Application in Rehabilitation Motion Analysis

Shayan Shokri[1], Shane Ward[2], Pierre-Amaury M. Anton[3], Paolo Siffredi[1], Guglielmo Papetti[1]

[1]Polytechnic University of Milan, [2]Pennsylvania State University, [3]Swiss Federal Institute of Technology Lausanne (EPFL)

*Abstract*—The increase in world elderly population has significantly underlined the need for continuous health care measurement, specifically in rehabilitation monitoring. The new technologies has enabled people to have in home healthcare services, meanwhile, motion analysis methods are widely used for human activity monitoring as a remote healthcare service. Wearable sensors have indicated promising results both in convenience and technical performance. These sensors are extensively used in human motion analysis and advancement of wireless communications has intensively contributed to this field. Exploiting wireless technology and wearable sensors contributes to more effective help in emergency cases and has significantly decreased the hospitalization time. This paper reviews the most recent advances in wearable sensors used in motion analysis, specifically in the field of rehabilitation. Firstly, common wearable sensor technologies are introduced and then wearable sensors deploying Carbon Nano Tubes (CNT) are specifically reviewed. The next section is dedicated to sensor fusion in which possibility and performance of integration of new technologies are reviewed. This technique has been widely exploited to bring forth certainty in clinical results. Lastly, the challenges and future possibilities for advancement in motion analysis sensors is discussed.

## I. INTRODUCTION

In recent years, rapid advancements have been made in wearable technology. Meanwhile, health industry has one of the highest growth rates in exploiting these sensors. The desire for more accurate human activity monitoring has driven vast progress. Wearable sensors allow for monitoring outside of a clinical setting into a patient's daily life. This allows for more personalized, consistent, and immediate health evaluation.

Wearable technologies for human activity monitoring have commonly been applied to rehabilitation motion analysis. A number of maladies and diseases underline the need for motion analysis from Parkinson's disease to strokes to accidents causing trauma which affects motion ability. Using sensors embedded into wearable technology, numerous motion evaluation variables could be monitored including, balance, stability and other gait variabilities. The field of rehabilitation motion analysis has seen immense growth over recent years.

## II. WEARABLE SENSORS FOR ACTIVITY MONITORING

Wearable sensors are electronic devices which are suitable both to detect important physiological parameters and to be worn by a person. They can be embedded into several pieces of clothing such as watches, shirts, belts, shoes, or directly positioned on the body. To be employed in such a way, wearable systems need to be easily worn, without encumbering any activity of the subject and they need to be wireless. Wearable sensors can be widely exploited for the ubiquitous and remote monitoring of patients who have undergone surgery, as well as patients who need to follow a rehabilitation process or elderly at risk of falls, reducing the pressure on health-care systems and the related costs. Telehealth applications may allow patients for a continuous monitoring, supporting independent living, which is perceived as an essential feature especially for elderly. A wearable monitoring system may comprise several wearable sensors, which provide information about a few physiological parameters. The raw data has to be collected and eventually be processed by other components and algorithms in order to be finally monitored. The improvement of MEMS (Micro Electro-Mechanical Systems) technology has enabled the development of a wide scope of faster and more efficient sensors [1, 2]. One of the most employed sensors for activity monitoring are accelerometers, which are able to measure acceleration along specific axes and over a given range of frequencies. Although accelerometers can rest on different methods of transduction (e.g. piezo-resistance or variable capacitance), they are all based upon the same principle of a suspended mass which causes a spring to compress or stretch proportionally to the acceleration they are submitted. Gyroscopes represent another type of commonly used sensors, which can measure angular velocity by means of a vibrating mass which is displaced when the device rotates. Furthermore, magnetometers can be employed to detect the orientation of the user's body in relation to the Earth's magnetic field, which is aligned with a sensitive axis of the device [1, 3].

Nowadays, the occurrence of falls has been rising together with the increasing number of elderly people. The consequences of fall events are often detrimental, since they may include hospitalization, morbidity and in most cases, a reduced quality of life. According to the World Health Organization, falls are the second leading cause of unintentional injury deaths, mostly among adult older than 65 years of age [4]. Therefore, an accurate detection of falls is important to provide prompt assistance and to avoid the so-called "long lie" situations, which have shown to increase the mortality rate.

Several solutions, either wearable or non-wearable, have been presented to develop an appropriate fall detection system. Non-wearable solutions rely on ambient sensors installed within the living environment. On the other hand, the wearable approach typically employs cheaper MEMS sensors, which are used for a ubiquitous monitoring of elderly as well as to provide an automatic detection of falls. Montanini et al. have recently developed a fall detection system based on a pair of sensorized insoles, which are provided with a tri-axial accelerometer and three force resistive sensors each. These resistors are based on a variable resistance which changes according to the applied force, and they have the purpose of monitoring the body weight distribution on the sole of the foot and the walking activity. A processing unit for each insole gathers the data provided by the force sensors and the accelerometer, allowing to analyze the gait cycle, to detect potential falls, and to transmit the collected information as well. The accelerometer, the processing unit and a battery are hosted in a suitable cavity in the shoe sole. The system has shown a good accuracy, low computational costs, and a low energy consumption; furthermore, the user's needs have been fulfilled, due to a minimal obtrusiveness which allows to walk comfortably without any bother caused by the presence of the device [5].

A shoe-based system has been presented also by Sazonov et al. [6], but the aim of this wearable device was to perform posture monitoring. The assessment of patients' posture by means of wearable devices has several applications in different areas of rehabilitation, since it may enable an independent training as well as tele-rehabilitation. The system proposed by the authors is lightweight and unobtrusive, and it is able both to provide insole pressure monitoring and to identify many static postures. The sensing components are five force-sensitive resistors, positioned under critical points of contact on the insole, and a 3-axis MEMS accelerometer placed on the back of the shoe. The developed system is able to provide both a posture analysis and an estimation of the energy expenditure [6]. The monitoring of postural sway is important for patients with neurological disorders or head injuries; in laboratory settings, postural sway is commonly measured by means of force plates which detect the Centre of Pressure of the ground reaction force. A research, reported on [7], has shown that Inertial Measurement Units (IMUs) have an even higher sensitivity and accuracy as compared to the most common laboratory-based systems, proving that they could be efficiently employed in user-friendly devices to monitor and analyze posture of patients outside of clinical environments. A Smart Rehabilitation Garment has been proposed by Wang et al. [8]; this wearable system has the aim of both supporting a correct posture of the patient's trunk and notifying any compensatory movement that may occur, providing a physical feedback to the person by means of a vibration. Accurate measurements of the posture are allowed by the employment of two IMUs and an Arduino processor which reads the data obtained by the sensors and provide information about any bending of the upper trunk. The garment is designed to be as comfortable as possible, allowing for a long-term use. Another important activity monitoring that wearable sensors may perform is human gait analysis. A clinical assessment of gait can be useful to detect motor disorders and to support diagnosis and personalized rehabilitation processes. The traditional and expensive camera-based methods for gait assessment do not allow an appropriate analysis in non-structured environments outside of a laboratory. Hence, IMU-based wearable systems have gained a growing interest in the last few years, since they provide reliable measurements of limb motion in daily life conditions, without any constraints of light or space. The portable gait analysis system developed by Qiu et al. [9] consists of up to eight lightweight IMUs for each side of the lower limbs, a microcontroller, and a wireless communication module. This system is able to identify the different phases of a gait cycle, due to the data provided by the gyroscope, and to assess several gait parameters, such as the stride length and cadence, the gait speed, and gait symmetry, which represents an important feature in rehabilitation processes. These parameters have to be properly evaluated since they may change significantly in stroke patients, individuals with neurological disorders and elderly. A different wearable system for gait analysis has been, finally, developed by Bae et al. [10]; the tele-monitoring system proposed by the authors is made up of an IMU attached to the forefoot of a shoe provided with ground reaction force sensors. The information gathered by the sensors are then transmitted via internet to a remote location, where a therapist might monitor it. This system enables the monitoring of the patients' gait status during daily living activities and outside of rehabilitation facilities, allowing for a customized treatment.

## III. WEARABLE SYSTEMS IN ACTIVITY MONITORING

The great development of sensor-based technologies and MEMS in the last years have led to great achievements in the wearable-devices field for medical applications. In particular, wearable systems for human motion analysis can provide objective and immediate symptom monitoring of neurological conditions such as Parkinson's disease (PD) and stroke. Approaches based on accelerometry have been widely exploited [11], and the small size and the low weight of the sensors have promoted their use [16]. This is particularly important because these systems can provide helpful motion parameters to assess the effectiveness of physiotherapy rehabilitation exercises and can reduce evaluation time.

Concerning Parkinson's disease, whose patients are affected by motor disorders, wearable motion devices are suitable for gait analysis. A couple of inertial sensors, for example placed on shoes, can unobtrusively quantify in detail meaningful parameters such as cadence, step extension and step velocity in order to objectively track and value symptoms of Parkinson's disease like brief steps, walking instability and tremor. Therefore, data collected allow to get gait patterns, increase assistance to diagnosis, keep under observation disease advancement and assess response to treatment [12].

Alterations of gait are typical elements of walk in Parkinson's disease patients and freezing of gait is a common condition, especially as the disease proceeds. Freezing of gait is a sudden, complex and variable episode of inability to perform the next step, usually upon gait beginning or turning rapidly. This event may cause falls and consequently injuries, less independence and fear of possible other falls. Freezing of gait episodes can be automatically detected using a triaxial accelerometer-based device, worn in the pockets or on a belt, which will accurately measure frequencies in the acceleration signals and compare pathological frequencies to normal walking ones, that are lower. Small wearable sensors provide practical, objective and reliable identification and quantification of freezing of gait events, supporting specialists and experts in clinical evaluation [13].

Video or direct observations and clinical scales are considered the gold standard in clinical motion analysis and are used to validate parameters provided by wearable devices [11,12].

Stroke survivors are often affected by motor disabilities. Rehabilitation process, requiring accurate motion analysis, is based on the exact determination of motor deficit, the establishment of adequate treatment and a following progress evaluation. Wearable sensors offer the possibility to assess patients' capacity, integrate the subjective observation of an expert and get data for a longer period of time. That allows characterization of movements in daily-life conditions even outside clinical setting, increase the patient's motivation and promote personalized rehabilitation exercises.

However, motion analysis in stroke patients is very challenging because their movements are often very slow and segmented in relation to the ones of healthy people [14].

Stroke survivors and people who have suffered from neurological injuries may show reduced activity of paretic limbs and motor impairment. Sensors like Inertial measurement units (IMUs) may distinctly measure movements and angular displacement while performing activities, allowing to analyze limbs functionality. An IMU-based device, comfortably worn on the wrist, can assess arm movements in terms of duration and intensity while, for example, a person is performing simple tasks like reaching and handling an object in everyday situations. Movements cause angular variations between the arm and horizontal and sagittal plane, that are detected by the sensor to rate the limb positioning and then compared to the healthy arm. Moreover, these measurements could be important parameters comparable to clinical tests commonly used in patients with motor impairment, without the influence of ambulatory exercises as a small wearable device may permit evaluation in the home environment [15].

Combining IMUs with electromyography sensors (EMG) may also contribute to more complete motion analysis by the study of activity and functionality of muscles in the detection of movements, gestures and motor patterns [14].

Wearable devices for activity monitoring are increasing interest in the field of medical applications. In addition to movement and gait analysis already described for stroke and Parkinson's disease patients, wearable systems have been used to evaluate the outcome of orthopedic surgery, for example by assessing hip or knee extension and flexion after replacement surgery or ligament reconstruction surgery. Furthermore, providing auditory feedback or tactile feedback (through vibrations, preferable in outside environments), it is possible to correct gait in order to decrease joint loads, improve walking stability and reduce high oscillating movements, with applications in osteoarthritis and in people with vestibular loss [16].

Nowadays, a wide range of wearable devices for activity monitoring, whose functions can also be integrated into smartphones, is available. They provide data and parameters not only usable in the clinical setting, but also exploitable to improve and promote walking and exercises. This is important both in patients with motor disorders and in people who need to increase their physical activity levels. Indeed, physical activity is very important and related to the prevention and management of many conditions. It helps maintain weight and provide health benefits for the elderly, people after orthopedic surgery, physical therapy programs and generally stimulate change a sedentary behavior. Easy pedometers or more advanced and accurate multi-sensor systems are able to measure activity, in particular energy consumption, time span, intensity and frequency of exercise. Feedback from the device and goal setting can motivate to fulfill everyday recommended physical activity, grow self-efficacy and provide a personalized management of physical activity in terms of quantity and quality. Wearables should enable real-time activity monitoring while performing everyday routines or physical exercise and allow remotely assessment and recommendations about the effectuation of exercise from experts [17].

However, despite the growing potential and interest in wearable devices, there are challenges. Patients may have difficulties related to the interaction with technology and devices should be small, comfortable, easy to wear and to keep in the proper position in order to improve confidence with their usage [11]. Moreover, accurately detecting slow movements is very challenging, because they provide very low changes of signal [14], as well as distinguishing voluntary movements from passive involuntary ones [11].

Wearable devices have the great advantage of being able to be used for prolonged gait analysis not confined to medical laboratories [16]. Another important aspect could be the ability to detect motor symptoms even at the early stage of the disease when they are not evident and difficult to capture even by experts [12].

Although wearable devices for activity monitoring are becoming increasingly present in several studies for clinical applications, it is fundamental to carefully consider advantages and disadvantages of possible usages and, above all, a solid validation of the devices relying on standard references [11].

IV. SENSORS

A. Electromyography Sensors

Movement of the body is generated from muscles contractions and expansions. These muscle movements result from neurological signals sent from the cerebral cortex to the necessary nerve ending [18]. An electromyography (EMG) sensor measures the electrical signals given off by the muscles during contraction. Most commonly two forms of EMG are used: intramuscular EMG and surface EMG (sEMG), a noninvasive method. Intramuscular EMG is an invasive method which utilizes a small needle to enter into the muscle. Alternatively, surface EMG is noninvasive, instead using an electrode to measure a signal which propagates to the surface of the skin. sEMG is preferred to use for information about timing or intensity, however, does not provide as accurate measurements as a intramuscular EMG would [19]. Study of muscle contractions can allow for early and accurate detection and classification of motion [18,20].

A disadvantage of EMG sensors is the necessity for electrode wires. These wires transmit the signal from the muscle surface to the sensing unit. This increases the obstructiveness of the device for patient use. Obstruction can cause huge difficulties for patients when the wearable sensor is intended for everyday use. Obstructiveness is often minimized by small electrode wires with a sensing element attached to the limb being measured [21].

EMG signals require a high level of processing for analyzing and classifying due to their complicated form. Their high sensitivity to noise adds increased complications for analysis. Noise is generated inherently in EMGs as well as by electromagnetic radiation, electrocardiographic effects, or cross talk from muscle groups different from the muscle of interest. Motion applications become especially noisy due to movement artifacts. Motion artifacts are unavoidable due to the nature of muscle lengthening creating movement of skin and electrode relative to one another. Figure 1 shows the complexity of raw data collected from an EMG. These complexities require algorithms for processing and converting results into accurate and understandable data [19].

The future of EMG sensing lies in the ability to create low computational cost algorithms for accurate manipulation of this data. The complexities of these algorithms for lower limb human motion are being studied and applied for EMG fusion with other sensors [18].

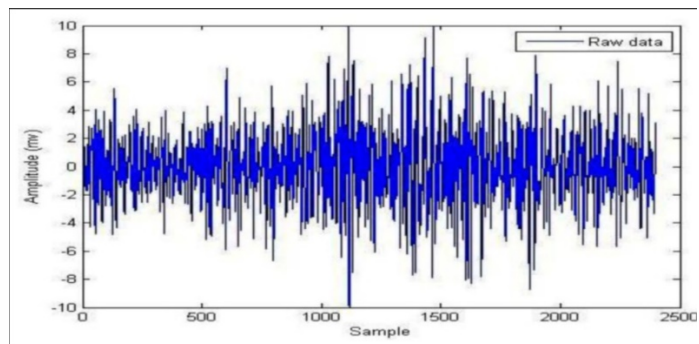

Fig. 1. Raw EMG signal [19].

### B. Sock-Type Electromyography Sensors

Recovering muscle activity after a fall is largely correlated to the lower leg muscle activity and multiple factors of that activation. Despite the importance of measuring this muscle activity, EMG sensors have been difficult to implement into every day wearable devices. The nature of these devices makes this difficult for multiple reasons. Most predominantly, EMGs are hard to implement in everyday use because of the technical skill required for usage. A patient unfamiliar in anatomy would have difficulty placing the electrodes correctly everyday for measurement of the desired muscles [22].

A new application of EMG sensors has been studied to effectively remove the barriers to everyday patient use. A research group has proposed the idea of a sock-type EMG for daily use. Several lower leg muscles of particular interest to motion propagate their signals to around the ankle. Electrodes placed at this region would be capable of measuring these distal signals to evaluate muscle contractions in regions of interest higher up. The placement of the electrodes in a garment like a sock remove the need for patient knowledge in anatomy to use [22].

To prevent significant motion artifacts by large relative displacements between the electrode and skin, garment style electrodes require constant compression of the electrode to the skin. The elastic band of socks allows the electrode to maintain its flush position against the lower leg to minimize signal and allow a constant measurement [22].

### C. GNSS Receiver

The position of the device, its velocity and precise time (PVT) is obtained from a GNSS-receiver. The Global Navigation Satellite Systems combine global systems : GPS (United States) , GLONASS (Russia), Galileo (Europe), BDS (China), and regional systems : QZSS (Japan), IRNSS (India). The satellites send signals containing its position in space obtained from the almanac which contains the data of all the GNSS satellites (e.g. present and future location) and the date and time of the emission. The GNSS-receiver captures those signals from at least four satellites. Three satellites are necessary for the trilateration to calculate the user's latitude, longitude and altitude. The receiver does not dispose of an atomic clock versus to satellites. Therefore the fourth satellite is used to provide the precise time to the device to ensure an accurate positioning. The receiver uses usually more than four satellites to avoid any interferences in order to increase the precision. Moreover, the main interferences are due to physical factors (e.g. reflection on buildings, forests) and atmospheric factors (troposphere and ionosphere). One of the solutions consists in algorithms of correction for the atmospheric interferences. Differential GPS augment as well the precision of localization with fixes receivers on earth that are in contact with the device. The Wide Area Augmentation System (WAAS) applies this principle for a larger area and results in a localization with a precision of less than 2 meters with an accuracy of 95%.[23] However, WAAS is only available in North America. Besides, similar systems have been developed in Europe (EGNOS), in India (GAGAN) and Japan (MSAS). Moreover, in 2017, the first dual frequency GNSS receivers has been presented for smartphones. A dual-frequency GNSS receiver provides a significant improvement in the error correction by capturing more than one frequency [24].

### D. Liquid-state Conductive based Stretchable *Sensors*

To detect the mechanical deformation, strain sensors have been widely employed. Meanwhile, to make these sensors exploitable in biomedical applications, stretchability is still one of the main remaining challenges. To be more specific the new technology should be capable of higher stretchability while maintaining the high linearity and high sensitivity. Regarding to this challenge, Functional electrodes patterns of strain sensors with arbitrary deformability design has been studied [25-27]. Meanwhile, Graphene, carbon nanotubes, metal nanoparticles and nanowire have been excessively studied and it is understood that these materials are intrinsically subject to deformability limitation.

Accordingly, new studies have been carry out on Carbon nanotubes (CNT) and silicone rubber (SR). These two materials are used together to reach theoretically infinite stretching feature along with high electric sensitivity. To scrutinize performance of this structure, Herein, a liquid-state conductive based upon CNTs and SR has been exploited for strain sensors and revealed remarkable features. The gauge factor (GF=(ΔR/R0)/(ΔL/L0)) for this developed strain sensor is 43.84 while reaching remarkable 400% elongation ratio.

Accordingly, this fully-stretchable strain sensor would be a promising candidate in biomedical applications. The proposed highly-sensitive strain sensor has been deployed in human joint The gauge factor (GF=(ΔR/R0)/(ΔL/L0)) for this developed strain sensor is 43.84 while reaching remarkable 400% elongation ratio [28].

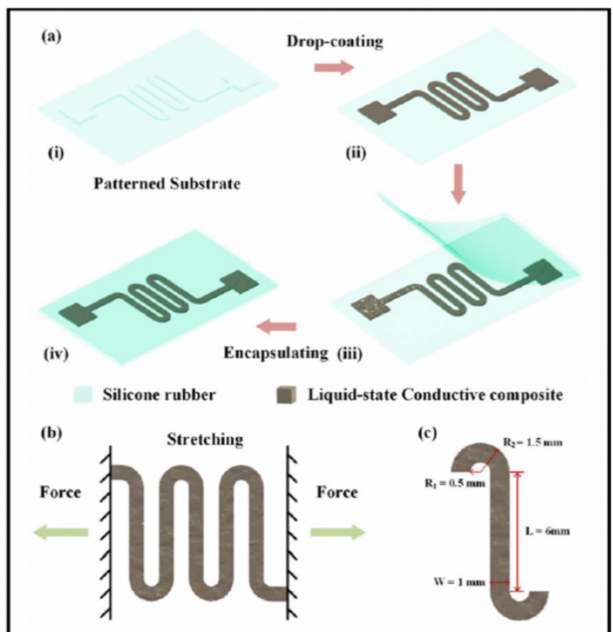

Fig. 2. (a) highly-sensitive fully-stretchable sensor design process based on liquid-state conductive composite. (b) Schematic of the serpentine structure, (c) a unit cell dimensions [28].

Accordingly, this fully-stretchable strain sensor would be a promising candidate in biomedical applications. The proposed highly-sensitive strain sensor has been deployed in human joint measurement and due to its remarkable linearity, significant stability and decent hysteresis performance, it has indicated promising results. It is worth mentioning that using silicone rubber, the strain sensor has demonstrated to be significantly stable enduring 360 degree twisting and 4-folded state. This performance is shown in Fig. 3. As it has been demonstrated the silicone rubber size is comparable to size of a typical coin and this enhances the employability of this technology in accurate and delicate cases. Applications include spinal analysis and finger bending measurement.

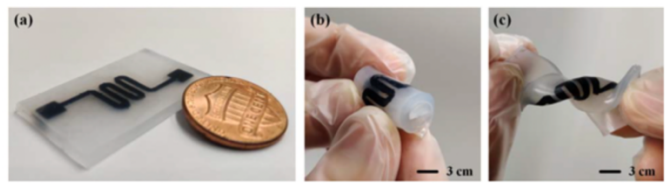

Fig. 3. Mechanical features of highly-sensitive fully-stretchable strain sensor. (a) photograph comparing the coin size with highly-sensitive fully-stretchable sensor, (b,c) the sensor indicates remarkable stretchability and flexibility: (b) 4-folded, (c) 360 degree twisted [28].

**E. Conductive Hydrogel Sensor**

In recent years flexible wearable hydrogel technology has undergone extensive advancement. The main features include, lightweight, biocompatibility, high sensitivity, potential self-healing capability, adhesive ability and robust stretchability. These characteristics have contributed to new researches in exploiting these flexible hydrogels in human activity monitoring [29]. External Pressure, strain, heat, light and voice stimuli are precisely transferable to electric signals using this technology [30]. Obtaining accurate electric signals, polyaniline, polypyrrole, polythiophene, graphene and carbon nanotubes (CNT) are implemented in hydrogel matrixes, physically and chemically. Among these, CNTs are widely used for electrical connectivity and conductive nanocomposite preparation. Considering stimuli-responsive feature and electrical connectivity, as key features needed in human motion analysis, N-isopropylacrylamide (NIPAM), a thermal sensitive monomer, would be a fitting candidate to be exploited along with CNT structures [31]. To prolong life span of these structures, host–guest interaction, dynamic chemical bonds, H bonds, metal coordination complex are employed to enable self-healing features [32].

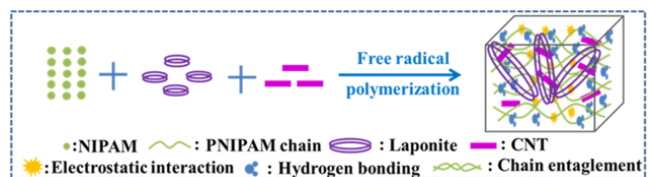

Fig. 4. FT-IR spectra of laponite and hydrogels [32].

This technology has extensive capability to be exploited in biomedical applications. Concerning human activity monitoring, wearable hydrogels are used for physical environment recognition with comparatively accurate voice, heat and light stimuli response. Besides, they are employed in synovial joint muscles activity due to pressure and strain sensitivity. The new features of this CNT-based technology enables clinicians to use it in more complex motion scenarios. Hence, this would accurately recognize the surface through reflected sound waves and could be extensively useful to help elderly in emergency cases activity estimation.

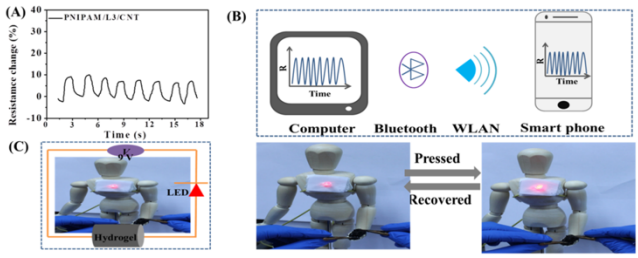

Fig. 5. (a) Relative resistance change of the sensor based on varying pressure (b) The diagram indicating schematic of wireless transmission process for real-time resistance monitoring, (c) Indicating sensor of hydrogel on a mannequin to test pressure-dependent conductivity by LED bulb [32].

### F. Textile Knit Stretch Sensor

Considering clothing as the second layer of skin, the idea of utilizing intelligent textiles to monitor human activity has been extensively studied. This allows human to use this sensors conveniently and exploit biomedical measurement features without mounting on any extra device. Textile knit stretch sensor employs commercial knits as resistive conductive fabrics. [33]

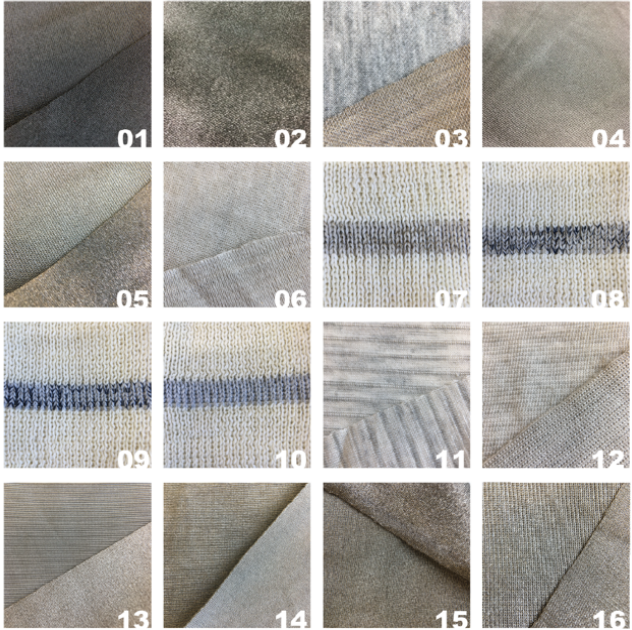

Fig. 6. Photographs of sixteen sample textile sensor [33].

Analyzing different materials used in commercial e-textiles, the gauge factor was observed ranging from 0.87 to 2.97. Meanwhile, silver-plated yarn, single bed jersey knits indicated high linearity while having low stretchability and a narrow working range. Therefore it can be exploited in tracking muscle, back, and spine. On the other hand, EeonTex is knitted silver-plated yarn and knitted spun stainless steel yarn indicates the least hysteresis and fastest responsiveness, hence, this sensor is a strong and suitable candidate for rapid, dynamic movement tracking such as knee angel measurement. The most valuable advantage of this sensor is the low production cost while maintaining a high resolution. It is applicable in biomedical measurements specifically in knee bending angel rehabilitation measurements, although in numerous other dynamic human motion detection cases this sensor is exploitable.

## V. SENSOR FUSION

Due to the complexity of motion, numerous variables affect gait (speed, stability, quality, symmetry) [39]. Each of these variables allows for a brief glimpse into the state of motion or balance. Reliance on many variables for complete motion analysis can create imprecisions in single sensor measurements. The integration of many of these variables into one interface allows potential for a more complete understanding of the current evaluaion of human motion in a patient. Use of sensor fusion can involve multiple measurements for a more complete and accurate analysis of motion states [21].

### A. Accelerometer-Gyroscope-Magnetometer Fusion

Accelerometers, gyroscopes, and magnetometers are commonly fused to form an inertial measurement unit (IMU). State-of-the-art inertial measurement units use microelectromechanical systems technologies to allow 9-axis measurements. This means each sensor is measured independently along the x, y, and z axis. Accelerometers measure linear acceleration. Ascertaining this data is based on measuring the direction and magnitude of the g-force which in MEMS technology varies with linear motion. Gyroscopes measure angular velocity based on the Coriolis effect. This uses a mass and spinning disc with an axis of rotation free to take any direction. Magnetometers orient the motion with respect to Earth's magnetic field [39].

The accelerometer has a bad signal to noise ratio at the onset of acceleration. Gyroscopes alternatively measure instantaneous measures because the axis of rotation is not affected by instantaneous tilting because of the conservation of angular momentum. Yet gyroscopes' accuracy breaks down over time increasing error. When cooperatively fused, the accuracy of the motion is greatly improved. Accuracy becomes independent of time as both are used for their respective advantages. The magnetometer then adds a common reference frame to add accuracy of orientation for a more comprehensive understanding of the motion [39].

### B. IMU-Force Fusion

The fusion of IMU and force sensors allows the integration of multiple variables of motion analysis for more accurate representation of fall risk and other gait abnormality. IMU detects, among other variables, body sway through linear acceleration, angular velocity, and direction of movements. This is an important variable for determination of balance and fall risk. These can be placed at the trunk, head, and lower limbs of the patient for complementary fusion of multiple IMUs [39]. Plantar force sensors can quantify postural stability and gait variability by measuring center-of-pressure (COP) and stance/swing time [21,39]. The placement of force sensors on the plantar surface, a region imperative in both static and dynamic balance controls, increases understanding of the patients ability to support balance at the time of detection [21].

Increased lateral COP has previously been proven to be an indicator of deterioration of static stability. Similarly, COP

variability can be indicative of worsening dynamics balance [39]. Each of these can present a fall risk to the patient. Together with the IMU data, fall risk assessment, as well as other important motion parameters, greatly increases its accuracy.

When fused with IMU sensors, force sensors help improve detection and classification of many gait defects. Another variable, abnormal swing time, can suggest risks such as a Freezing of Gait (FOG) episode in Parkinson's patients [20]. Often when the input information is determined to be hazardous to the patient a biofeedback system is implemented to alert the patient at risk [21]. These systems have been made compatible with smart phones for easy use by patients. The real time alerts help prevent many accidents in patients susceptible to gait abnormalities. These systems allow real time correction of a patient's motion of which they otherwise would have been unaware of. In one study implementing a similar biofeedback system for an IMU-force sensor, all patients saw significant improvement in both static and dynamic stability [39].

**C. IMU-EMG Fusion**

The combined use of IMU and EMG sensors is currently a popular technique for motion analysis [21]. Signals from these independent measurements of varying natures greatly expand the extent and accuracy of information collected. EMGs' measurements of motion inducing muscle activation is more responsive creating faster measurements to changes. However, patients often cannot intuitively decipher raw EMG data to acknowledge harmful patterns. IMU however supply more intuitive data for motion such as acceleration and velocity rather than muscle activation [18]. Together the distinct signals have been proven to become more reliable than either sensor individually [20].

The fusion technique has been commonly and successfully applied in many upper limb motion studies. However, the diverse characteristics and complications of lower limb motion create large computational costs in the needed algorithms for fusion. These costs should be limited for real time analysis as well as to decrease equipment needed for EMGs that can already be obstructive. Therefore, uses for lower motion analysis are state-of-the-art and largely experimental [18]. A recent study has successfully applied machine learning algorithms to determine between four ambulation states (jump, walk, descend, ascend) [18].

Another study applied similar algorithms based on IMU-EMG fusion to determine between voluntary and involuntary movement, an indicator of dangerous gait patterns [40]. This study was used to recognize tremor and dyskinesia, two forms of involuntary movement. Each of these presents a danger to the patient which can be mitigated by early detection and warning.

Other studies have been able to achieve determination of more specific motion pathophysiology. One such example accomplished accurate judgement between various subtypes of FOG episodes [20]. One subtype, shuffling forward, is characterized by short, shuffling steps with little forward movement of the feet. The other, trembling in place, features lack of postural adaption and an immediate inclination of the trunk. Each subtype is caused by different neurological dysfunctions, so distinction between them allows more accurate diagnosis. EMGs were placed at the tibialis anterior and gastrocnemius muscles in the lower limbs. Early or increased activation of these muscles has been proven to indicate a FOG episode [20].

Collection of this data by EMG and IMU can help beyond immediate improvement of patient motion. The underlying pathophysiology of these disorders are not yet fully understood. Collection of data from both the EMG and IMU can help researchers improve scientific understanding of these conditions as well as improve the safety and lives of patients affected by it [20].

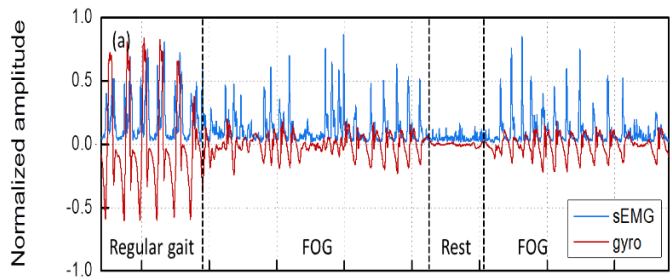

Fig. 7. Fusion resulting in combined IMU and EMG data. It indicates periods of rest, regular function, and gait dysfunction [20].

**D. Use of IMU for Fall Detection**

Falls constitute a serious risk of injuries for elderly and may sometimes lead to death if no assistance is rapidly provided.[34]

The detections and alerts must be the most accurate possible to avoid any dissatisfaction from the user. Too much fall alerts can become frustrating for seniors. In fact, it is important to create a real distinction between the activities of daily living and falls. In the case the user sits down quickly no fall alarm must be reported.

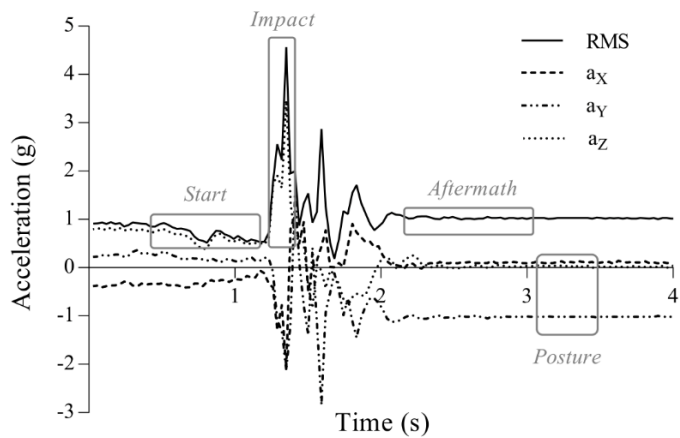

Fig. 8. Acceleration changes during an accident fall [35].

As a matter of fact, it's possible to identify some typical acceleration changes during a fall as shown on the figure 1. The Root Mean square of the accelerations on each axes of the accelerometer (x, y and z) in our case is equal to :

$$RMS = \sqrt{\frac{1}{3}(a_x^2 + a_y^2 + a_z^2)}$$

When the subject starts falling down on the floor, the acceleration is characterized by a value lower than 1 g; the impact with the ground can be identified by a peak in the acceleration signal; after the accident, the subject usually remains motionless for a period of time (Aftermath phase) and the measured acceleration is around 1 g; finally, the orientation of the body can be detected. The wearable fall detection system proposed by Pierleoni et al. contains a MARG (Magnetic, Angular Rate, Gravity) sensor and it is based on the detection of the phases previously mentioned.[35] The MARG sensor is composed of an 9-axis IMU and an orientation filter (e.g. Kalman-Filter). The raw data collected by the system's sensors, composed of a 3-axis MEMS accelerometer, a 3-axis MEMS gyroscope and a 3-axis MEMS magnetometer, are fused together by a Micro Controller Unit which is able to provide a comprehensive measurement of body's orientation, to detect potential falls, and to manage alarm signals. This fall detection system has been improved by adding a barometer to the previously described sensors [36]. The barometric sensor is exploited to measure the altitude variation in a fall, during which an increase of the atmospheric pressure acting on the device occurs. Since the signal provided by the barometer is affected by a great amount of noise, a fusion with the vertical acceleration data obtained by the accelerometer is necessary to have a reliable altitude estimation. The system has shown a high accuracy in the detection of different type of falls and a good acceptance by the users, since it has been designed as a waist-mounted device.

Algorithms have also an evident influence on the fall detection. Recently, the extreme learning algorithms (ELM) have confirmed superior results: up to 98.65% of accuracy, 92.10% of sensitivity and 97.48% of specificity [41].

**E. IMU and GNSS Receiver Coupling**

Since the 2000's, GPS positioning has made some serious improves in precision and democratization. The technology that at the beginning was only used for military purpose is now accessible with a remarkable precision for all the population.

The application of GPS technology with AI-based algorithms enables fast and reliable results, specifically in trajectory scenarios with moving objects, which could be exploited in real-time applications [37, 38].

The fusion of the IMU and the GNSS-receiver leads to a more precise analysis of the user's motion. This fusion can also be relevant in the case of fall detection. If the person is unconscious or in the inability to call for help, the device can call the emergency services and give its actual position.

This feature is present on fall detection bracelets. Nowadays, it has been also implemented in smartwatches (e.g. Apple Watch Series 4 and 5 ). It has recently proved that it can save lives by shortening the time of arrival of the emergency services [42].

## VI. CONCLUSION

Recent advances in innovative technologies such as MEMS have led to a growing interest in wearable sensors for activity monitoring.

They provide objective evaluation of motion parameters and the possibility to monitor activities remotely and continuously, even outside clinical environments fostering home-based rehabilitation. Their small size allows to track movements unobtrusively and should not cause any discomfort to the people who wear them.

Wearable devices have been used to monitor activity in Parkinson's disease patients and in stroke patients, allowing to accurately quantify symptoms and motor functions, and supporting experts in clinical evaluation. Feedback and goal setting may help to correct gait patterns and increase motivation in their usage.

Most of the sensors used in wearable devices for activity monitoring are accelerometers, often fused with gyroscopes and magnetometers to form a common IMU. Indeed, sensor fusion can provide a more complete and accurate motion analysis.

However, despite the growing development of wearable systems, some challenges remain. It is important to promote more user-friendly wearable devices, which may be embedded in lightweight and comfortable garments in order to make people more willing towards these solutions. To enable a prolonged monitoring of patients, wearable systems should also rely on long-lasting batteries with lower consumption.

Further developments are also needed to make wearable devices suitable for a wider range of applications, allowing for an accurate measurement of slow movements, which are an important feature for elderly or patients who have to follow rehabilitation processes.

In conclusion, further research is essential to ensure a stable usage of wearable devices for activity monitoring in health care and rehabilitation applications.